\documentclass[11pt,twoside]{article}
\usepackage{asp2006}
\usepackage{psfig}
\usepackage{epsf}
\usepackage{graphicx}
\usepackage{lscape}
\def\edcomment#1{\iffalse\marginpar{\raggedright\sl#1\/}\else\relax\fi}
\reversemarginpar

\begin{document}

\def\kms{$\mathrm {km s}^{-1}$}
\def\msun{\rm M_{\odot}}
\def\mbh{M_{\rm BH}}
\def\etal{{et al.\ }}
\def\ergs{\rm erg\,s^{-1}}
\def\bfJ{\bf {J}}
\def\simlt{\mathrel{\rlap{\lower 3pt\hbox{$\sim$}}\raise 2.0pt\hbox{$<$}}}
\def\simgt{\mathrel{\rlap{\lower 3pt\hbox{$\sim$}} \raise 2.0pt\hbox{$>$}}}
\def\lsim{\mathrel{\rlap{\lower 3pt\hbox{$\sim$}}\raise 2.0pt\hbox{$<$}}}
\def\gsim{\mathrel{\rlap{\lower 3pt\hbox{$\sim$}} \raise 2.0pt\hbox{$>$}}}
\def\di{\mbox{d}}
\def\msunpc3{\msun~{\rm {pc^{-3}}}}
\newcommand{\be}{\begin{equation}}
\newcommand{\ee}{\end{equation}}

\title{A path to radio--loudness through gas--poor galaxy mergers and
  the role of retrograde accretion} \author{M. Dotti,$^1$ M. Colpi,
  $^2$ L. Maraschi,$^3$ A. Perego,$^4$ M. Volonteri$^5$} \affil{ $^1$
  Max Planck Institute for Astrophysics, Karl-Schwarzschild-Str. 1,
  85741 Garching, Germany\\ 
  $^2$Dipartimento di Fisica, Universit\`a degli Studi di
  Milano-Bicocca, Piazza Della Scienza 3, 20126 Milano, Italy\\
  $^{3}$ INAF, Osservatorio Astronomico di Brera, Via Brera 28, 20123 
  Milano, Italy\\
  $^4$ Department of Physics, University of Basel, Klingerbergstr. 82, 4046 Basel, Switzerland\\
  $^{5}$Department of Astronomy, University of Michigan, Ann Arbor, MI
  48109, USA }

\begin{abstract}
In this note, we explore a pathway to radio--loudness under the
hypothesis that {\it retrograde} accretion onto giant ($M_{\rm BH}\sim
10^9 \, \msun$) spinning ($a\gsim 0.7$) black holes (BHs)
leads to the launch of powerful jets, as seen in radio--loud QSOs and
recently in LAT/Fermi and BAT/Swift Blazars.  Counter--rotation of the
accretion disc relative to the BH spin is here associated to {\it
  gas--poor} galaxy mergers progenitors of giant (missing--light)
ellipticals.  The occurrence of retrograde accretion enters as
unifying element that may account for the radio--loudness/galaxy 
morphology dichotomy observed in AGN.
\end{abstract}

\vspace{-0.5cm}
\section{Relativistic jets and the radio-loud morphology dichotmy}

Thanks to a wealth of observational and theoretical studies of AGN
from radio to gamma-ray frequencies, radio-loud AGN are presently
thought to contain relativistic jets which propagate from the nucleus
out to kpc and, in some cases, Mpc scales.  The power of jets required
to at least fuel their radio lobes is huge, up to $10^{47}\ergs$
corresponding to an energy release of $\sim 10^{62}$ erg over a time
of $10^8$ yr.  In radio-quiet AGN, the radio emission is instead weak
and large scale radio jets are absent.  Although the boundary between
these two classes is blurred, strong jets (in terms of kinetic
relative to accretion power) exist only in a minority of AGN.

Optical studies of {\it nearby} galaxy nuclei (Capetti \& Balmaverde
2006) have shown that {\it all} radio-loud AGN invariably reside in
giant {\it core} (or {\it missing--light}; Kormendy et al. 2009)
ellipticals, i.e. in early-type galaxies whose nuclear surface
brightness profile shows a {\it deficit} in star-light with respect to
the outer profile.  By contrast, normal, less massive {\it coreless}
(or {\it extra--light}; Kormendy et al. 2009) ellipticals {\it all}
host radio-quiet AGN (Capetti \& Balmaverde 2006), suggesting that a
{\it morphology-related dichotomy} exists among {\it ellipticals},
likely determined by galaxy evolution. The dichotomy seems also to
emerge from a sample of radio-loud and radio-quiet QSOs at redshift
$z<0.2$, the former residing in giant ellipticals that show signs of
interaction, and the latter residing in merging gas-rich galaxies of
intermediate mass (Wolf \& Sheinis 2008). Evidence exists that the
most powerful radio and gamma--loud AGN are associated to the heaviest
supermassive BHs in the universe, with $M_{\rm BH}\gsim 10^9\,\msun$
(Ghisellini et al. 2009a,b) indicating that the BH {\it mass} can be a
key parameter for radio-loudness.

From a theoretical point of view, the launching of jets from
accreting BHs have been extensively studied in the past three decades,
following the seminal paper by Blandford and Znajek (BZ hereafter;
1978).  The BZ process exploits energy extraction from a rapidly
rotating black hole via a purely electromagnetic interaction of a
large scale magnetic field which threads the
rotating event horizon. In this framework, jets are produced at the
expense of the rotational energy of the black hole $E_{\rm rot}=M_{\rm
  BH}[1-2^{-1/2}(1+\sqrt{1-a^2})^{1/2}]c^2$, where $a$ is the
dimensionless spin parameter related to the BH spin ${\bf J_{\rm
    BH}}={\bf {\hat l}}\,(GM^2_{\rm BH}a/c)$ pointing along ${\bf
  {\hat l}}$ (for a maximally rotating [$a=1$] Kerr BH of
$10^9\,\msun$, $E_{\rm rot}=6\times 10^{62}$ erg).  This may suffice
to power a jet and to explain why AGN with comparable optical luminosities can be either radio-loud (large $a$) or
radio-quiet (small $a$, as $E_{\rm rot}$ decays $\propto\,M_{\rm BH}a^2/8$
for $a\to 0$) depending on $a$ (Wilson \& Colbert 1995; Sikora et al. 2007).

Advanced fully general-relativistic magneto-hydrodynamical simulations of spinning BHs 
(McKinney 2005, 2006) predict 
high collimation of the inner Poynting flow, 
bulk Lorentz factors of order 10, and jet powers $P_{\rm jet}\propto a^2$, or
$\propto a^6$,  implying in this last case high jet luminosities only for close to maximally spinning
BHs (Tchekhovskoy et al. 2009). Another essential ingredient determining the BZ power is the value
of the magnetic field threading the horizon, which is tied to the accretion rate and disc structure
(Ghisellini et al. 2009a). 

In two recent works, Garofalo (2009a,b) stressed the importance of the
plunging region between the innermost stable circular orbit
(ISCO\footnote{ISCO, expressed in units $GM_{\rm BH}/c^2$, is at 6 for
  $a=0$, and at 9 (1) for retrograde accretion with $a=-1$ (prograde
  accretion with $a=1$).}) of the accretion disc and the BH horizon.
In this region the magnetic field can be substantially amplified from
its value at ISCO, particularly if the BH has a {\it
  retrograde} spin vector with respect to the accretion disc angular momentum (we will use 
  here the convention of negative spin, $a<0$, for retrograde accretion). As a 
  consequence, high jet powers can be extracted
even/also from non-maximally rotating BHs (Garofalo 2009b).

Here, we explore the consequences of Garofalo's conjecture
in the aim at connecting the BH spin and accretion mode to the
structure of the underlying host galaxy. Moderately high values of the
spin parameter $a$ ($\gsim 0.7$) and retrograde accretion appear to enter as
unifying ingredients to account for the radio-loud/host morphology dichotomy observed, at low redshifts, among
ellipticals.\footnote{See Sikora et al. (2007) for the discussion on
the most general late/early-type versus radio-quiet/loud dichotomy observed in AGN, and
Fanidakis et al. (2009) for a different spin dependent cosmological model of jet formation in AGN.}
These considerations go in the direction of the very recent proposal that most radio-loud AGN harbor
spinning BHs accreting in the retrograde mode (Garofalo et al. 2010).
 
\section{A pathway to radio--loudness through dry galaxy mergers}

Can retrograde accretion onto a spinning BH 
be established during galaxy evolution?

\subsection{Isolated disc galaxies}

To answer this question in a broader context, we first consider {\it
  isolated disc} galaxies, in which BHs are expected to grow {\it
  only} by accretion. In these galaxies BH fueling can occur either
via multiple uncorrelated episodes of accretion (Moderski et
al. 1998), or via secular bar-in-bar instabilities (Shlosman,
Frank \& Begelman 1989) that remove gradually the gas angular
momentum.  
In the first mode, episodes of chaotic
accretion result in low BH spins (King \& Pringle 2006; Volonteri et
al. 2007). Prograde and retrograde accretion change both the BH spin
modulus $a$ and direction ${\bf{\hat l}}.$ Since retrograde orbits
carry a larger angular momentum than prograde orbits, random episodes
result in average values of $\langle a\rangle \lsim 0.4$.  In this
scenario, retrograde accretion events are common, but the accreting BH
is always slowly spinning, so no powerful, long--lived collimated jets
are produced. In the second scenario, the BH grows through {\it coherent}
accretion. Even if $a\sim 0$ initially, accretion can spin the BH up
to $a\sim 1$ after doubling its mass (precisely, a factor
$\sqrt{6}$, Bardeen 1970). In this scenario, the BH spin ${\bf J}_{\rm
  BH}$ {\it aligns} with the angular momentum of the accretion
disc. The BH remains in the radio-quiet mode if $a\lsim 0.9$, as
retrograde accretion is unlikely to establish, given the coherence of
the flow. A strong jet  in a disc galaxy could be triggered by prograde accretion only 
if $a>0.9$ (Tchekhovskoy et al. 2009), but the jet power would be
$\sim 30$ times dimmer than that produced by retrograde accretion onto a
maximally spinning BH (Garofalo 2009b).

\subsection{Galaxy Mergers}

The situation is different in the case of {\it major mergers}, i.e.
mergers between galaxies
of comparable masses.
Galaxy mergers are commonly divided in two classes, gas rich (or {\it
  wet}) mergers where the fraction of
cold gas is large ($\gsim 10\% $ of the mass in stars), and gas poor
(or {\it dry}) mergers where gas is a
small fraction of the stellar mass and has no dynamical effect in the
merger.

\subsubsection{Wet mergers}

During a {\it wet} merger, the tidal field between the two disc galaxies
drives large amounts of gas (up to $50 \%$ of the total gas mass)
toward the centres of the two interacting galaxies (Mayer et
al. 2007).  When the two nuclei later merge in a single structure, the
gas settles into a dense, selfgravitating circumnuclear disc in which
the BH relative orbit becomes {\it circular} and {\it corotating}
until the BHs form a Keplerian binary (Dotti et al.  2009). In this
phase, lasting $\lsim 10^7$ yr, the BHs accrete
in a coherent manner at a rate sufficient to {\it align} their spins,
initially oriented at random, to the angular momentum of the nuclear
disc (Liu 2004; Bogdanovic, Reynolds \& Miller 2007; Dotti et al.  2010): in
response to the Bardeen--Petterson warping of the small--scale
accretion discs grown around each BH, total angular momentum
conservation imposes fast ($\lsim 1$ Myr) alignment of the BH spins
with the angular momentum of their orbit and so of the large--scale
circumnuclear disc (Dotti et al. 2010). Thereafter, accretion remains
prograde until coalescence, with no major changes in the BH spin
orientation.  Under these circumstances the BH remnant retains the
spin direction of the parent BH spins, both oriented parallel to the
angular momentum of their orbit: the post-coalescence BH may thus
acquire a large spin $a \gsim 0.7$ (Berti \& Volonteri 2008; Kesden, Sperhake \& Berti 2010), sum of
the internal and orbital spins. Because of {\it fast alignment} and
spin coherence relative to the flow, {\it prograde} accretion
continues even after BH coalescence and so {\it no powerful
  long--lived jets are produced, according to our working hypothesis,
  unless the BH remnant is close to be maximally rotating.}

We can also follow the evolution of a wet merger by looking at the
properties of the galaxy remnant. Gas rich galaxy mergers are expected
to form either spheroids with residual rotation or discs, depending on
their initial gas content (Robertson et al. 2006), encounter geometry
and the presence of cold gaseous streams flowing onto the evolving remnant
(Governato et al. 2009). The stellar nucleus of the relic galaxy can
in principle be perturbed by the massive BH binary.  The binary
hardens ejecting stars via three-body scatterings (Merritt \&
Milosavljievic 2004) creating a {\it stellar core}.  The
post-coalescence BH can further heat the stellar nucleus if it
experiences a strong gravitational--wave induced recoil ($\lsim 1000
\,$ km s$^{-1}$; Lousto \& Zlochower 2009 and references therein): in
its oscillations back to the galactic centre the BH deposits its
kinetic energy expanding further the core (Boylan--Kolchin et
al. 2004; Gualandris \& Merritt 2008). However, in a gas--rich merger,
these effects are strongly suppressed: the stellar deficit by the BH
binary is weak since the binary hardens mainly via gaseous torques
(see Colpi \& Dotti 2009 for a review) and {\it extra--light} is produced following the
formation of new stars inside the circumnuclear disc (Hopkins et
al. 2009a).  Furthermore, spin-orbit alignment from coherent accretion
implies small gravitational kicks for the relic BH, limiting the
effect of dynamical heating of the nucleus.  As a consequence, {\it we
  predict that wet mergers result mostly in radio--{\bf quiet} AGN hosted by
  coreless, extra--light elliptical or spiral galaxies}.  Figure 1
illustrates schematically the evolution of the two BHs and of the
remnant galaxy in a gas--rich merger.

\begin{figure}
\begin{center}
\includegraphics[scale=0.45] {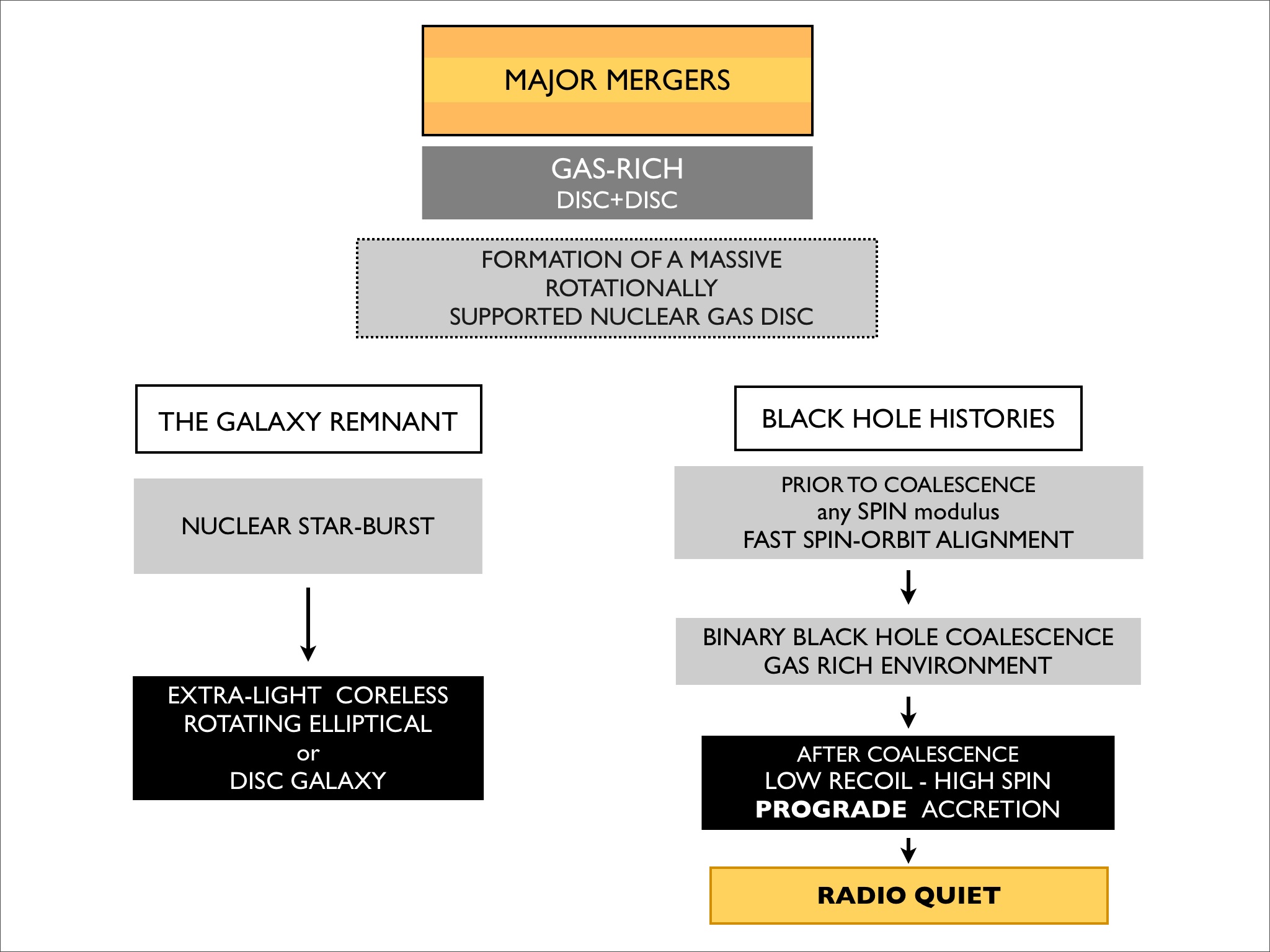}
\end{center}
\caption
{Gas--rich mergers between nearly equal mass disc galaxies. The left
  and right branches of the scheme refer to the evolution of the
  galaxies and the BHs, respectively. Left: We hypothesize that the
  merger ends with the formation of a normal coreless elliptical where
  the central extra--light results for a major star formation episode
  triggered inside the massive circumnuclear disc.  Galaxy evolution
  models do not exclude the survival or re--growth of the disc
  (Governato et al. 2009) so that a disc galaxy may form at the end of
  the merger.  Right: Observations of normal coreless ellipticals
  indicate that they host BHs with $M_{\rm BH}<10^8\,\msun$ (Kormendy
  \& Bender 2009).  The BHs that initially inhabit the two galaxies
  are expected to pair and form a binary {\it corotating} with the
  disc (Dotti et al. 2009).  During inspiral under the action of
  gas-dynamical torques, gas accretion aligns the BH spins to the
  orbit (Dotti et al. 2010) so that the post-coalescence BH has a high
  ($a\gsim 0.7$) spin and {\it corotates} with the disc. After this
  phase and according to our working hypothesis the BH is {\it
    radio-quiet}. We notice that a short-lived radio-loud phase could
  be present, prior to BH coalescence, before disc-orbit and
  spin-orbit alignment occur.  }

\end{figure}

\subsubsection{Dry mergers}

Mergers of {\it gas--poor} galaxies (e.g. between two ellipticals)
lack massive nuclear gas discs.  The BHs thus complete their
inspiral in a collisionless background of stars.
The BH spins are expected to be uncorrelated with the geometry of the
encounter, so that, as the binary forms, their spins are randomly
oriented relative to the binary orbit.  The BHs may capture gas 
and experience episodes of
retrograde accretion until they merge. At coalescence the BH spin
changes in modulus and orientation, and memory is lost of the initial spin
directions.  Berti \& Volonteri (2008) predict, for isotropic dry
mergers, a broad distribution of final spins centred around $\langle
a \rangle \sim 0.7$ (close to the value resulting from the coalescence
of two non--spinning BHs) and also large recoil speeds (e.g. Lousto et al.
2009).

The end--result of a dry merger is expected to be a {\it giant, missing--light}
elliptical  whose nuclear region has been
shaped by the BH (Hopkins et al. 2009b; Kormendy \& Bender 2009). Hardening through three--body scatterings with stars
excavates a {\it stellar core} in the nuclear region, further expanded
by BH heating by the recoil. The final giant rapidly spinning BH later settles in the
galaxy's centre and starts accreting.  In the absence of a rotationally
supported nuclear gas structure, half of the accretion events will be
retrograde, and the BH enters the {\it radio-loud}  phase.  

If the BH is very massive ($M_{\rm BH}>10^9\,\msun$), as seen in giant
ellipticals (Kormendy \& Bender 2009), we argue that its
spin direction remains {\it stable} through all accretion episodes: 
 because of its large BH mass and the lack of a massive nuclear disc, the BH 
angular momentum likely exceeds the angular momentum carried 
by the accretion disc (King et al. 2005).
Indeed, we note that for giant BHs, the warp radius $R_{\rm warp}$ (delimiting 
the distance for gravito-magnetic interaction between the
  BH and the disc) exceeds the outer radius of the accretion disc determined by
self-gravity, $R_{\rm out}$.  Following Perego et al. (2009), for a
  $10^9\,\msun$ BH, the outer radius $R_{\rm out}\sim 40\, M_{\rm
    BH,9}^{-52/45} f^{-22/45}_{\rm E} R_{\rm G}$ (where $R_{\rm G}$ is
  the Schwarzshild radius and $f_{\rm E}$ the Eddington factor) is
  smaller than $R_{\rm warp}\sim 10^3\,a^{4/7} M_{\rm BH,9}^{4/35}f_{\rm
    E}^{6/35} R_{\rm G}$ implying that over the Bardeen-Petterson timescale 
    ($\lsim 100$ yrs) the {\it entire} disc {\it aligns} o
  {\it antialigns} (depending on the disc's initial orientation) with
  the BH spin vector that keeps its orientation stable.
    Thus the BH can sustain a collimated jet
over large scales until accretion ceases.  

{\it We thus expect
  radio--loud AGN to be hosted in missing--light, core ellipticals,
  remnants of dry mergers}. This second scenario is summarized in
Figure 2.

\begin{figure}
\begin{center}
\includegraphics[scale=0.45]{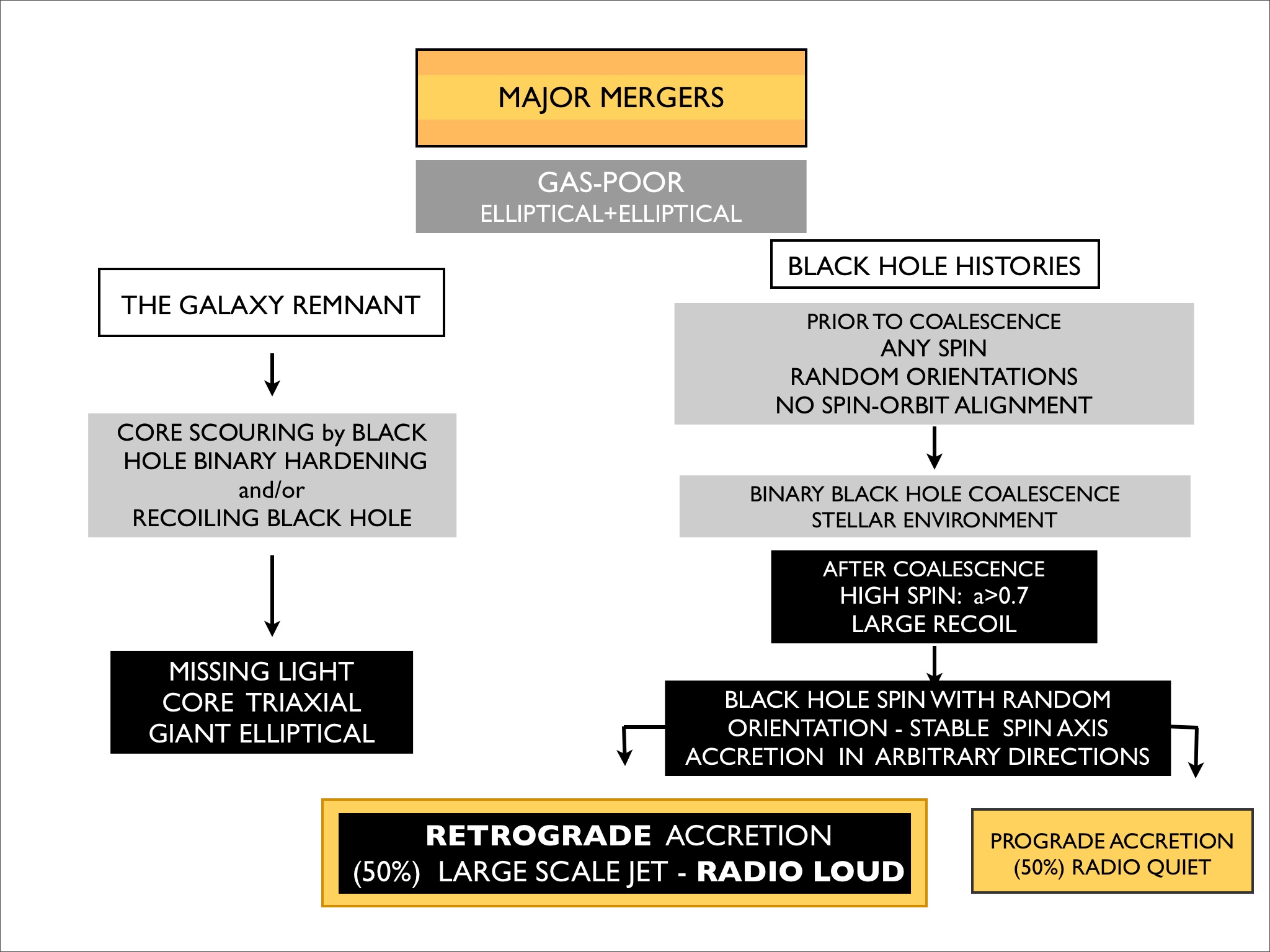}
\caption{
Gas--poor mergers between nearly equal mass galaxies (e.g. between two ellipticals). The left and right branches of the
scheme refer to the evolution of the galaxies and the BHs,
respectively. Left: We hypothesize that the merger ends with the formation of a giant missing--light 
elliptical where a stellar core has been excavated by the BH binary in its hardening by
scattering off single stars, and/or by heating from gravitational recoil after BH coalescence.
Observations of giant core ellipticals indicate that they host very massive BHs with $M_{\rm BH}\gsim 10^9\,\msun$
(Kormendy \& Bender 2009). 
Right:  The BHs that initially inhabit the two galaxies are expected to pair, form a binary and
eventually coalesce under the action of stellar torques. Prior to coalescence the two BHs
have arbitrary spin moduli and random orientation.  In
the gas--poor environment the spins of the giant BHs do not align with the orbit and
the post-coalescence BH ends with a high spin $a\gsim 0.7$ and large recoil.
Once settled at the centre of the galaxy host the BH may experience episodes of
{\it retrograde} accretion (in 50\% of the cases) and become, according to our hypothesis a 
{\it radio--loud} AGN. In the remaining 50\% cases the BH may be active and become
an AGN. Observations indeed indicate that giant ellipticals host radio--quiet AGN, and this
is a natural outcome of our working model. }
\end{center}
\end{figure}

\section{Discussion}

In this note, we propose that {\it retrograde} accretion onto a
massive ($M_{\rm BH}\gsim 10^9\,\msun$) {\it post-coalescence}
spinning BH is conducive to the generation of a powerful large-scale
jet, in the gas-poor environment of a giant elliptical with core.
This is in agreement with the observation that radio--loud AGN are
hosted in bright ellipticals that show a stellar core (whenever
optical data are available),  and with the recent finding on the
  dichotomy in radio jet orientations in elliptical galaxies (Browne
  \& Battye 2010).  In less-louder radio-loud AGN, Browne and Battye
  find a tendency for the axis of the radio emission to align with the
  minor axis of the starlight of the host, an oblate rotationally
  supported elliptical.  By contrast they find no preferred
  radio-optical alignment among the radio-louder objects possibly
  hosted in triaxial non-rotating ellipticals.  This study appears to
  support our evolutionary scheme: coreless (extra--light)
  rotationally supported ellipticals (Kormendy et al. 2009) that
  likely form in gas-rich mergers are expected to host a BH accreting
  in the prograde mode with a high spin aligned with the angular
  momentum of the large-scale disc, and so aligned with the minor axis
  of the starlight. By contrast, core (missing--light) triaxial
  ellipticals are expected to host a BH that has a random spin
  orientation regardless its accretion mode (i.e. whether prograde or
  retrograde).


The model rests on Garofalo's conjecture that around a retrograde BH,
magnetic fields threading the horizon reach maximum amplification,
resulting in a dramatic enhancement of the jet power even for
non--extreme values of $a$.  The interval of spin values for operation
of this process is still weakly constrained by theory. Furthermore
{\it prograde} accretion can also result in large BZ luminosities (Garofalo
2009a,b). Thus, there should exist a {\it zone of avoidance} delimited
by a maximum--negative $a^{-}_{\rm jet}$ and a minimum--positive
$a^+_{\rm jet}$, outside which the generation of jets is possible.
The values of $a^-_{\rm jet}$ and $a^+_{\rm jet}$ may not be symmetric
as non--symmetric is the underlying Kerr spacetime.  If retrograde
accretion is a necessary condition {\it only} for the most powerful
jets hosted in ellipticals, the value of $a^-_{\rm jet}$ that our
model requires is $a^-_{\rm jet}\gsim -0.7$, resulting from isotropic
BH coalescences in dry mergers.  If prograde accretion vehicles the
production of jets as well, a value of $a^+_{\rm jet}$ close to $+0.4$
(from chaotic accretion) would imply the presence of jets in any type
of galaxy, from discs in isolation to spheroids in clusters and
groups. The paucity of radio--loud disc hosts and the lack, at low
redshift, of radio--loud coreless host--ellipticals suggest that such
a low value of $a^+_{\rm jet}$ is irrealistic.  One would be tempted
to require $a^+_{\rm jet}\sim +1$, according to the models by McKinney
(2005).

The rotational energy stored in giant BH ($M_{\rm BH}\gsim 10^9$) may
suffice to power a radio--loud AGN at an average level of $\sim
10^{46}\,\ergs$ for a lifetime $\tau_{\rm jet}\sim 10^8$ yr.  
However retrograde accretion causes the BH to spin down
on a time $\tau_{\rm down}\sim$ min($\tau_{\rm ac},\tau_{\rm jet}$).
Spin--down by accretion occurs on $\tau_{\rm ac}\sim J_{\rm BH}/({\dot
  M_{\rm net}{\tilde l}_{\rm ISCO}}) \sim M_{\rm BH}/{\dot M}_{\rm
  net}$ (where ${\dot M}_{\rm net}$ is the {\it net} inflow rate onto
the BH, and ${\tilde l}_{\rm ISCO}$ the angular momentum per unit mass
of a test particle at ISCO).  $\tau_{\rm ac}$ is close to the Salpeter time for
$e$--folding of the BH mass.  The BZ
timescale $\tau_{\rm jet}$
depends on the magnitude of the magnetic field that threads
the horizon and on the thermal/geometrical structure of the accretion disc
(Moderski et al. 1998). In Garofalo's scenario, retrograde accretion
is a transitory phase during the lifetime of a BH (Garofalo et al. 2010) whose
duration and recurrence depend on the way the BH is fed whether through continuous or
chaotic  accretion.

Recent observations of extremely bright Blazars with BAT/{\it Swift}
at redshifts $z>2$ (Ajello et al. 2009; Ghisellini et al. 2009a) and of
Blazars and FRGs with LAT/{\it Fermi} (Ghisellini et al. 2009b) indicate 
that luminous radio-loud AGN host {\it giant} BHs,  and 
that a prominent accretion disc co-exists with a powerful jet suggesting that extraction
of rotational energy occurs through accretion (Maraschi 2001). 
We are tempted to associate the extreme BH mass, inferred from the
disc luminosity, to the stability of the BH spin orientation and the jet power to
its spin modulus and to accretion in the retrograde mode.  Further studies on the {\it
  stability} of retrograde accretion will help in disentagling the
nature on the AGN radio--loud/quiet dichotomy.

\vskip 1.2 truecm

\acknowledgements 

M.V. acknowledges support from NASA award ATP NNX10AC84G. \\

\noindent
{\bf References}

\vskip 18pt
\noindent
Bardeen J.M., 1970, Nature, 226, 64\\
Berti E. \& Volonteri M., 2008, ApJ, 684, 822\\
Blanfdord R.D. \& Znajek R.L., 1977, MNRAS, 179, 433\\
Bogdanovic T., Reynolds C.S. \& Miller M.C., 2007, ApJ, 661, L147\\
Boylan--Kolchin M., Ma C.P. \& Quataert E., 2004, ApJ, 613, L37\\
Browne I.W.A. \& Battye R.A., 2010, aXiv:1001.1409\\
Capetti A. \& Balmaverde B., 2006, A\&A, 453, 27\\
Colpi M. \& Dotti M., 2009, Advanced Sci. Lett. in press (arXiv:0906.4339)\\
Dotti M., et al.,  2009, MNRAS, 396,1640\\
Dotti M., et al., 2010, MNRAS, 402, 682\\
Fanidakis N. et al., 2009 submitted to MNRAS, arXiv:0911.1128\\
Garofalo D., 2009a, ApJ, 699, 400\\
Garofalo D., 2009b, ApJ, 699, L52\\
Garofalo D., Evans D.A. \& Sambruna R.M., 2010, MNRAS, arXiv:1004.1166v1\\
Ghisellini G. et al., 2009a, to appear in MNRAS, arXiv:0909.0932\\
Ghisellini G. et al., 2009b, submitted to MNRAS, arXiv:0912.0001\\
Governato F. et al., 2009, MNRAS, 398, 312\\
Gualandris A. \& Merritt D., 2008, ApJ, 678, 780\\
Hopkins P.F. et al., 2009a, ApJS, 181, 135\\
Hopkins P.F. et al., 2009b, ApJS, 181, 486\\
Kesden M., Sperhale U. \& Berti E., 2010, submitted to PRD, arXiv:1002.2643\\
King A.R., Lubow S.H., Ogilvie G.I. \& Pringle J.E, 2005, MNRAS, 363, 49\\
King A.R. \& Pringle J.E, 2006, MNRAS, 373, L90\\
Kormendy J., Fisher, D.B., Cornell M.E. \& Bender R., 2009, ApJS, 182, 216\\
Kormendy J. \& Bender R., 2009, ApJL, 691, 142\\
Liu F.K., 2004, MNRAS, 347, 1357\\
Lousto C.O. \& Zlochower Y., 2009, Phys RevD, 79, 4018\\
Maraschi L., 2001, AIP Vol. 586, Eds. Wheeler \& Martel, p. 409\\
Mayer L., et al., 2007, Science, 316, 1874\\
McKinney J., 2005, ApJ, 630, L5\\
McKinney J., 2006, MNRAS, 368, 1561\\
Merritt D. \& Milosavljevi\'c M., 2005, Living Reviews in Relativity, 8, 8\\
Moderski R., Sikora M., \& Lasota J.-P., 1998, MNRAS, 301, 142\\
Perego A., Dotti M., Colpi M. \& Volonteri M., 2009, MNRAS, 399, 2249\\ 
Robertson B., et al., 2006, ApJ, 645, 986\\
Shlosman I., Frank J. \&  Belegman M.C., 1989, Nature, 338, 45\\
Sikora M., Stawarz L. \& Lasota J.P., 2007, ApJ, 658, 815\\
Tchekhovskoy A., Narayan R. \& McKinney J., 2009, to appear in ApJ, arXiv:0911.2228\\
Volonteri M., Sikora M. \& Lasota J.-P., 2007, ApJ, 667, 704\\
Wilson A.S. \& Colbert E.J.M. 1995, ApJ, 438, 62\\
Wolf M.J. \& Sheinis A.I., 2008, AJ, 136, 1587\\

\end{document}